\documentclass{aastex}

\shorttitle{Luminosity Distributions of Cosmological GRBs}
\shortauthors{Kim, Chang, and Yi}

\begin{document}

\title{Effects of Luminosity Functions Induced by Relativistic Beaming
on Statistics of Cosmological Gamma-Ray Bursts}

\author{Chunglee Kim, Heon-Young Chang, and Insu Yi}

\affil{Korea Institute for Advanced Study, Seoul, 130-012, Korea}
\email{ciel@newton.kias.re.kr, hyc@kias.re.kr, iyi@kias.re.kr}

\begin{abstract}
Most of gamma-ray bursts (GRBs) models have predicted that the intrinsic
isotropic energy is limited to below $\sim 10^{53-54}$ergs.
Recently claimed high redshifts, correlation with supernovae,
and connection to the cosmic star formation activity point to a 
different energy requirement and source evolution history possibly with
strong beaming.
We study the effects of the beaming-induced luminosity function
on statistics of observed GRBs, assuming the cosmological scenario.
We select and divide the BATSE 4B data into 588 long bursts
(T$_{90}>2.5$~sec) and 149 short bursts (T$_{90}<2.5$~sec),
and compare the statistics calculated in each subgroup.
The $\langle V/V_{\rm max} \rangle$ of the long bursts
is $ 0.2901\pm 0.0113$, and that of the
short bursts is $0.4178\pm 0.0239$.
For luminosity function models, we consider a cylindrical-beam and
a conic-beam. We take into account
the spatial distribution of GRB sources as well.
A broad luminosity function is naturally produced when
one introduces beaming of GRBs.
We  calculate the maximum detectable redshift of GRBs,
$z_{\rm max}$.
The estimated $z_{\rm max}$ for the cylindrical-beam case is as high as
$\sim 14$ ($\alpha=1.0$) and $\sim 6$ ($\alpha=2.0$)
for the long bursts and $\sim 3$ ($\alpha=1.0$) and $\sim 1.6$ ($\alpha=2.0$)
for the short bursts where $\alpha$ is the photon index.
The large $z_{\rm max}$ value for the short bursts is rather surprising
in that the $\langle V/V_{\rm max} \rangle$ for
this subgroup is close to the so-called Euclidean value, 0.5.
We calculate the fraction of bursts whose  redshifts are larger than
a certain redshift ${\rm z'}$, i.e. $f_{\rm > z'}$.
When we take ${\rm z'}=3.42$ and apply the luminosity function
derived for the cylindrical-beam, the expected $f_{\rm > z'}$
is $\sim 75~ \%~ (\alpha=1.0)$ and $\sim 50~\%~ (\alpha=2.0)$ for long bursts.
When we increase the opening angle of the conic beam to
$\Delta \theta =3^\circ.0$,
$f_{\rm > z'}$ decreases to $\sim 20 ~\%~ (\alpha=1.0)$ at $ {\rm z'=3.42}$.
If we assume $\alpha=2.0$, the conic-beam with $\Delta \theta =3^\circ.0$
can not explain the redshift distribution of the observed GRBs.
We conclude that the beaming-induced luminosity functions are compatible
with the redshift distribution of the  observed GRBs although the apparent
"Euclidean" value of $\langle V/V_{\rm max}\rangle$ might be
explained by the standard model.

\end{abstract}

\keywords{cosmology: theory -- gamma rays: bursts -- methods:
statistical}

\section{Introduction}

The observation of high-redshift GRBs such as
GRB970508, GRB971214, and GRB980703
\citep{metzger97,kul98,dj98} confirmed
the cosmological origin of GRBs \citep{paczynski86,meegan92}.
High redshifts ($z>1$) of GRBs would imply a much larger apparent energy
budget than the conventional energy budget $\sim 10^{53-54}~$ergs unless
some significant beaming is involved \citep{ku99,kupi99}.
The isotropic energies of observed GRBs with
known redshifts range from $\sim 10^{50}~$ergs to $\sim 10^{54}~$ergs.
A broad luminosity function can be a reasonable answer for this wide
range, and can be naturally expected if we assume beaming of GRBs.
The suggested correlation between GRBs and
certain types of supernovae (Ib/c) implies that the total energy
available for GRBs could be rather limited at least in the low energy
GRBs $ L_{\rm int} \sim 10^{49}~$erg/sec
(Wang and Wheeler 1998; Woosley et al. 1999, see also Kippen et al.
1998; Graziani et al. 1999).
If we consider the supernova explosion as a universal GRB source,
then the reported GRB-supernova correlation implies
that the high redshift GRBs must be strongly beamed with the beam opening
solid angle, which is approximately given by
\begin{equation}
d \Omega \sim (L_{\rm int}/L_{\rm obs}) \sim
10^{-4} (L_{\rm int}/10^{49}{\rm erg/sec}) (L_{\rm obs}/10^{53} {\rm
erg/sec})^{-1}.
\end{equation}

When we consider beaming, a broad luminosity range and
a stringent constraint on the intrinsic energy are anticipated.
\citet{yi94} considered the effect of the luminosity function
due to the cylindrical-beam with the inhomogeneous distribution of GRB
sources on statistics of GRBs.
He analytically showed that the luminosity function follows a power
law with the index $-4/3$ in the very narrow beam model when $\alpha=1.0$.
His result has been confirmed by Mao and Yi (1994) and Chang and Yi (2000).
\citet{maoyi94} studied the statistical
properties of luminosity functions derived using
a conic beam with a uniform spatial distribution of GRB sources.
They examined the cumulative probability distribution of the peak
count rates. They found that the maximum redshift increases
as a product of the Lorentz factor and the opening angle decreases.
We attempt to generalize the luminosity function induced by the
cylindrical-beam by introducing the conical shape in such a beam.
If the progenitors of GRBs were produced in the late stage of
massive stars \citep{woo93,paczynski98,mac99},
their spatial distribution may follow
the star-formation rate (SFR) of massive stars
\citep{totani97,wijers98,bn00}.
We use the recently observed SFR data to derive the possible
number density distribution of GRB sources \citep{steidel99}.

In this paper, we study how beaming-induced luminosity functions
affect the statistics of the observed GRBs in the BATSE 4B catalog
\citep{batse4B} considering the two different spatial distributions
of GRB sources: an SFR-motivated distribution and the uniform
distribution.
We employ two different SFR-motivated distributions.
We discuss the effect of the photon index $\alpha$ ($=1.0$ and 2.0)
in the conlusion.
We assume a flat universe with no cosmological constant,
and adopt the Hubble constant $H_{\circ}=50{\rm km/sec/Mpc}$
(cf. Yi 1994; Mao and Yi 1994).
We analyze 775 observed bursts in the BATSE 4B catalog
in order to improve the statistical significance.
Assuming the bimodal distribution of durations of GRBs
\citep{ko93,lamb93,mao94,katz96} we divide the total sample
into two subgroups according to their duration
and study the statistics of each subgroup.

The selection criteria applied to observational data are
described in section 2.
Luminosity functions due to beaming are discussed in section 3.
We present results and discussions in section 4.
We summarise our conclusion in section 5.

\section{Classification of Observational Data}

We adopt the bursts in the BATSE 4B catalog \citep{batse4B} and calculate
their $\langle V/V_{\rm max} \rangle$ for selected GRBs \citep{schmidt88},
using fluxes in channels 2 and 3.
We choose bursts detected in 1024~ms trigger time scale with the
peak count rates satisfying $C_{\rm max}/C_{\rm min} \geq 1$.
Applying these criteria, we select 775 bursts among 915 bursts
in the BATSE 4B $C_{\rm MAX}/C_{\rm MIN}$ table.
The $\langle V/V_{\rm max} \rangle$ value for this total sample
is 0.3177 $\pm$ 0.0102.
Then we divide the selected sample into two subgroups
according to the burst durations ($T_{\rm 90}$),
which is motivated by the reported bimodal structure in GRB durations
\citep{ko93,lamb93,mao94,katz96}.
We define bursts with $T_{\rm 90} > 2.5~ {\rm sec}$ as
long bursts, and those with $T_{\rm 90} < 2.5~ {\rm sec}$ as short
bursts.
Their estimated $\langle V/V_{\rm max} \rangle$ values are
0.2901 $\pm$ 0.0113 (588 bursts) and 0.4178 $\pm$ 0.0239 (149 bursts),
for long bursts and short bursts respectively.
The number in parentheses indicates the number of bursts belonging to
each subgroup.
The $\langle V/V_{\rm max} \rangle$ values obtained from the two subgroups
show deviations from the values which are expected for a uniform
distribution in a Euclidean space.
In our subgroups, the long bursts have a larger deviation from the
Euclidean value. It agrees with the claim by Tavani (1998).
According to \citet{tavani98}, bursts with long durations ($T_{90}>2.5~{\rm
sec}$)
and hard spectra ($HR_{32}>3.0$) show the largest deviation
from the so-called Euclidean value.

\section{Beaming and Luminosity Functions}

We assume two kinds of number density distributions of GRB sources:
(i) the SFR-motivated distribution $ n_{\rm SFR}(z)$ and
(ii) the uniform distribution  $n_{\rm u}(z)=n_{\rm {\circ,u}}$.
We assume a two-sided Gaussian function as $n_{\rm SFR}(z)$ \citep{kim99},
and fit $n_{\rm SFR}(z)$ to the extinction-corrected SFR data
\citep{steidel99}.
The observed SFR gradually decreases or even keeps flat over a wide
redshift range beyond a certain redshift (cf. Madau et al. 1998).
The SFR-motivated number density distribution is approximated by
\begin{eqnarray}
n_{\rm SFR}(z)=& n_{\rm {\circ,SFR}} ~\exp \Bigl[ - {(z-z_{\rm c})^2
\over{\Delta z_1}^2} \Bigr],& (z < z_{\rm c}) \\
& n_{\rm {\circ,SFR}} ~\exp \Bigl[ - {(z-z_{\rm c})^2
\over{\Delta z_2}^2} \Bigr],& (z > z_{\rm c}), \nonumber
\end{eqnarray}
where $\Delta z_1$ and $\Delta z_2$ are widths of the fitting functions
for the left part and the right part and
$z_{\rm c}=1.5$ is a critical redshift.
All parameters are determined by fitting the observational
data. We set $\Delta z_1=1$ $(z < z_{\rm c})$ and
we also effectively fix $n_{\rm SFR}$(z) to $n_{\rm {\circ,SFR}}$
beyond $z_{\rm c}$.
The proportional constants
$n_{\rm {\circ,SFR}}$ and $n_{\rm {\circ,u}}$ are
normalized to satisfy the number from the  observational data.

\subsection{Cylindrical Beaming}

We consider the beaming-induced luminosity function following
\citet{yi93,yi94},
where the relativistic beam has a constant cross-sectional area, 
i.e. a perfectly collimated cylindrical beam.
Under this circumstance, a bulk motion of all the particles emitting
the photons is parallel to the cone axis. In this case,
the beam opening angle is irrelevant for the luminosity function.
We assume that the beamed
emission from "standard bursts" which have the same intrinsic
luminosity $L_{\rm int}$. The advantage of this model is that it physically
defines a range of the luminosity, i.e. $[L_{\rm min},L_{\rm max}]$,
for a given photon index $\alpha$, and the Lorentz factor $\gamma$.
An apparent luminosity function $\Phi(L)$ due to the uniformly distributed
beam in space is given by
\begin{equation}
\Phi(L) = {1 \over{p \beta \gamma}} ~L_{\rm int}^{1/p} ~L^{-(p+1)/p}
\end{equation}
where $\beta = (1-\gamma^{-2})^{1/2}$ and $p=\alpha+2$.
We adopt $\alpha=1.0$ \citep{mallozzi96} and 2.0 \citep{yi94}
and the Lorentz factor $\gamma=100$ (see Piran 1999).
If we assume a bi-polar jet, the minimum and the maximum luminosities
are naturally given as follows:
\begin{equation}
\L_{\rm min} =
L_{\rm int} B^p _{\rm min}, ~~ L_{\rm max}=L_{\rm int}B^p_{\rm max},
\end{equation}
where $B_{\rm min}=\gamma^{-1}$ and $B_{\rm max}=[\gamma(1-\beta)]^{-1}$
\citep{bland79}.
The intrinsic luminosity $L_{\rm int}$ for each subgroup is determined by the
$\langle V/V_{\rm max} \rangle$ test.
The obtained $L_{\rm int}$ values are $\sim 10^{43}~ {\rm erg/sec}$ for the
specific set of adopted beaming parameters.
This is a  much smaller value than that required in
the standard candle model, which is $L_{\circ} \sim 10^{51}~ {\rm erg/sec}$.

We calculate two statistical quantities using the beaming-induced
luminosity function explained in this subsection.
First, we calculate the $\langle V/V_{\rm max} \rangle$
as a function of threshold flux $F_{\rm th}$ \citep{che99}.
The threshold flux is obtained
when we put the maximum detectable redshift $z_{\rm max}$ in
\begin{equation}
F= {\Bigl( {{H^2_{\rm 0} L} \over{16 \pi c^2}} \Bigr)} ~
{{(1+z)^{1-\alpha}} \over{[(1+z)^{1/2} -1]^2}},
\end{equation}
where $L$ is the luminosity, $c$ is the speed of light, $H_0$ is the Hubble
constant, and $\alpha$ is the photon index.
After we obtain a set of $F_{\rm th}$, we calculate the
$\langle V/V_{\rm max} \rangle$ for each threshold flux.
The $\langle V/V_{\rm max} \rangle$ curve converges to 0.5
as $F_{\rm th}$ increases sufficiently large.
We convert flux to the number of bursts and plot
$\langle V/V_{\rm max} \rangle$
as a function of the number of bursts
for all bursts and long and short bursts in Fig. 1 (a) for $\alpha=1.0$ 
(thin lines) and $\alpha=2.0$ (thick lines).
The $\langle V/V_{\rm max} \rangle $ vs. number of bursts relation
essentially gives the same information that one obtains from the conventional
logN $-$ logP plot.
Our $\langle V/V_{\rm max} \rangle$ test does not include the threshold effect 
due to the detection efficiency.

Second, we calculate the fraction of GRBs.  We define the fraction of
bursts located at a redshift larger than $\rm {z'}$ as
\begin{equation}
f_{\rm > z'} = {{\rm {Number~of~bursts~with~a~redshift~larger~than}~z'}
\over {\rm {Total~number~of~bursts}}}.
\end{equation}
In Fig. 1 (b), we plot the fraction of bursts as a function of redshift
for each burst subsample for the cylindrical-beam case
with different number density distributions.

\subsection{Conic Beaming}

As a more realistic example, we adopt conically-beamed emission
from GRBs, which was studied by \citet{maoyi94}, and extended by
\citet{cy00}.
We consider the luminosity function produced using three different opening
angles,
i.e. $\Delta \theta = 0^\circ.1, 1^\circ.0$, and $3^\circ.0$.
We calculate $\langle V/V_{\rm max} \rangle$ and
the $f_{\rm > z'}$ for each subgroup with the derived luminosity
function.
The bulk Lorentz factor is set to be 100.
According to \citet{maoyi94},
the probability that we observe the bright bursts
rapidly increases as the opening angle increases,
while relatively dim bursts are not as detectable as brighter ones.
Therefore, as the opening angle increases,
the derived luminosity function becomes
similar to that of the standard candle case, in which
all bursts have the same maximum luminosity given
by this luminosity function.
When the opening angle is small
$(\Delta \theta \ll 1/ \gamma)$,
the luminosity function gives the same result as the cylindrical beaming
case.
Using this luminosity function,
we calculate the same statistics as we did in previous subsection.
 Results are shown in Figs. 2-5 for $\alpha=1.0$ (thin lines) and 
$\alpha=2.0$ (thick lines) in the cases of each sample.

\section{Results and Discussions}

The left panels in Figs.  1-5 (a) show the $\langle V/V_{\rm max}
\rangle$ as a function of the number of bursts.
We show results obtained from different luminosity distributions and
number density distributions. 
The thin lines in Fig. 1-4 represent the results obtained with 
$\alpha=1.0$ and thick lines $\alpha=2.0$.
 In Figs. 1-4, we assume
that the SFR-motivated number density distribution, which
is flat beyond a redshift larger than the critical redshift, $z_{\rm c}$.
Finally, we show our results obtained with the SFR-motivated
number density distribution, which gradually decreases
and $\alpha=1.0$ in Fig. 5 (a). In all cases, we assume that
n(z) rapidly increases if $z<z_{\rm c}$.
The dotted lines represent the observed $\langle V/V_{\rm max} \rangle$
curves with the $\pm 3 \sigma$ bound.
The solid lines indicate the theoretical results
derived from our luminosity functions with the SFR-motivated
number density distribution of GRB sources.
We plot the results obtained for the luminosity functions
with the uniform distribution of burst sources as dashed lines,
for comparaison.
Apparently all the luminosity functions we have studied
satisfy  the observed $\langle V/V_{\rm max} \rangle$
curve at the 3$\sigma$ significance level.
Although the number density distribution of GRB sources
is believed to correlate with the SFR, we cannot rule out
the uniform distribution of burst sources in this analysis
(see also Krumholz  et al. 1998).

The right panels in  Figs. 1-5 (b) show $f_{\rm > z'} $ in terms of $z'$.
The y-axis corresponds to the fraction of bursts which have redshifts
larger than a certain redshift $z'$.
Solid lines result from the SFR-motivated number density distribution
for the spatial distribution of GRBs and dashed lines the uniform
distribution.
Filled squares in  Figs. 1, 3, and 5 (b)
locate the known redshifts of 15 observed GRBs so far.
The redshift values are quoted  from
http://cossc.gsfc.nasa.gov/batse/counterparts/GRB\_table.html.
We compare the observed redshift values only with the long subsample 
because all GRBs with the known redshifts belong to the long bursts.
It is interesting to note that, if we assume $\alpha=1.0$,
a large fraction of GRBs are distributed at high redshifts,
provided that GRBs are assumed to follow the SFR, regardless of specific
beaming models. However, in the case of $\alpha=2.0$, $z_{\rm max}$
is significantly reduced, though it still remains quite large.
The fraction for $f_{\rm >z'}$ at $z'=3.42$
for a cylindrical-beam case is  $\sim 75~\%~$
for long and total samples when $\alpha=1.0$.
This is too large a value given the fact that the observed GRBs
with such high redshifts account for $\leq 10~\%$ of the total bursts.
This fraction become $< 50~\%$ when $\alpha=2.0$.
It is also interesting to note that,
though their $\langle V/V_{\rm max} \rangle$
value is close to a Euclidean value, most of the short bursts
are distributed at high redshifts in the case of $\alpha=1.0$.
When we assume that the long bursts
are uniformly distributed in space,
the $z_{\rm max}$ estimate for the broad beaming
cannot explain the largest observed redshift
in both cases of $\alpha=1.0$ and 2.0.

Our study is based on an assumption that
all GRBs have the same intrinsic luminosity and they are all beamed.
Because the $\langle V/V_{\rm max} \rangle$ test is not
sufficiently sensitive to the observed data,
different luminosity functions are essentially indistinguishable.
If there is an intrinsic luminosity function and
the degree of beaming of GRBs is
moderate, we may obtain results similar to what we have presented
in this study. If all GRBs are highly beamed and
their sources follow a SFR-like distribution, the maximum
detectable redshift becomes very large, even in the case that
the observed $\langle V/V_{\rm max} \rangle$ is sufficiently close
to the Euclidean value, i.e., $0.5$. Therefore, it is difficult to rule out
a possibility that the apparent Euclidean value of $\langle V/V_{\rm
max}\rangle$
may be due to the luminosity function, for instance, induced by beaming.
In other words, the so-called Euclidean value may have nothing to do with
the Euclidean distribution.
If the total burst sample is a mixture of broad (or nearly non-beamed)
beams and narrow beams, the exact fraction of the strongly beamed
bursts may affect the burst rate estimate significantly.
This is beyond the scope of this paper.

For a given Lorentz factor,
the range of the derived beamed luminosity distribution becomes broader
as the beam opening angle decreases.
We obtain $\log {L_{\rm max}/L_{\rm min}} \sim 13$ in the
cylindrical-beam case.
However, in the case of broad beaming,
$\log {L_{\rm max}/L_{\rm min}}$ becomes substantially smaller.
This is only an effective range because the cut-off for
the minimum luminosity given by the conic beam is rather loosely
constrained.
It is obvious that the luminosity range decreases
when the opening angle increases for a fixed Lorentz factor.
This is simply explained as follows:
Assuming that we observe a beamed emission with an opening
angle $\Delta \theta$, if we define $\psi$ as an angle between the
line of sight and the symmetry axis of the conic beam,
we can observe the beamed emission only within a certain range of $\psi$,
$\Delta \psi$. For instance, when $1/\gamma \ll  \Delta \theta$,
$\Delta \psi $ is approximately $\Delta \theta$.
For a given Lorentz factor, $\Delta \psi$ increases
as the opening angle of cone becomes larger.
The apparent luminosity range from
a broad beam would decrease gradually with $\psi$, and in the extreme case,
the isotropic radiation would have constant luminosity regardless of $\psi$.
However, when $\Delta \theta$ is very small, the apparent luminosity
rapidly decreases with $\psi$, and the derived luminosity range is
narrower than a broad-beam case.

\section{Conclusion}

We have demonstrated that the beaming-induced luminosity function may
explain the statistics of the  observed  GRBs.
In the case of the cylindrically beamed luminosity function,
we determined the intrinsic luminosity $L_{\rm int}$ by the
$\langle V/V_{\rm max} \rangle$ test for each subsample.
In all cases, the intrinsic luminosity
is much smaller than that of non-beamed luminosity functions, i.e.
$L \sim 10^{52-53}~ {\rm erg/sec}$.
The maximum luminosity obtained from our luminosity functions,
in both the cylindrical-beam and the conic-beam cases,
is $L_{\rm max} \sim 10^{50-51}~ {\rm erg/sec}$. This is
compatible with the required value for isotropic radiation.
The ratio between the maximum luminosity for the long bursts and
that for the short bursts $L_{\rm max,long}/L_{\rm max,short}$ is
$\sim 7$ for the narrow beam and $\sim 3$ for the broad beam, respectively.
This implies that these two subgroups may have
two different intrinsic luminosities,
and hence probably two different origin.
The obtained $z_{\rm max}$ in the cylindrical-beam case are
$\sim 10, 14, 3$ ($\alpha=1.0$) and $\sim 4.5, 5.5, 1.6$
($\alpha=2.0$)
for the total selected sample, and long and short subgroup, respectively.
The obtained $z_{\rm max}$ in the broader beam case
(i.e. $\Delta \theta=3^{\circ}.0$) are
$\sim 4, 6, 2$ ($\alpha=1.0$) and $\sim 2.2,2.6,1.2$ ($\alpha=2.0$)
for the  total selected sample, and long and short subgroup, respectively.
The maximum known redshift we consider, $z=3.42$, is estimated for GRB971214.
However, even a larger value of $z < 3.9$ is suggested for GRB980329
(http://cossc.gsfc.nasa.gov/batse/counterparts/GRB$\_$table.html).
Without any detailed information on the redshift distribution
of GRBs, the maximum redshift gives a rather strong constraint on
the ratio between the Lorentz factor and
the opening angle $\Delta \theta$.
For the conic beam case, the luminosity function derived for the narrow
opening angle seems to fit the $\langle V/V_{\rm max} \rangle$ curve
better. 

From this simple analysis we have carried out,
we have shown that the beaming-induced luminosity
function may account for the basic statistical properties of the observed
GRBs. The apparent Euclidean distribution of GRBs may be an indication of
the presence of the GRB luminosity function.

Although it is the beyond the scope of this work, it is interesting to point 
out that there could be a potentially observable correlation between
spectral hardness and luminosity since the most Doppler-boosted burst would 
probe the hardest part of the spectrum. Any wide scatters in intrinsic spectral
shapes would make the detection of this correlation complicated.

\acknowledgments
We thank H. Kim and  K. Kwak for useful discussions. IY is supported in part
by the KRF grant No. 1998-001-D00365.
We appreciate the referee, Ralph A. M. J. Wijers, for his suggestions and
some useful information.

\appendix

\clearpage

\figcaption{Two statistical properties of
three sample sets for the cylindrical-beam case are shown. Data sets
are indicated at the upper right corners in each panel.
In the left panels, we plot the $\langle V/V_{\rm max} \rangle$ curve
as a function of number of bursts.
Dotted lines represent
the observed $\langle V/V_{\rm max} \rangle$ with $\pm 3 \sigma$ bounds.
Solid lines and dashed lines show theoretical results, the former result
from the SFR-motivated number density distribution for the spatial
distribution of GRBs and the latter the uniform distribution.
In the right panels, we show
the $f_{\rm > z'}$ in terms of $\rm {z'}$ for each sample set.
The y-axis is the fraction of GRBs which would be
located farther than a certain redshift $\rm {z'}$.
Solid lines and dashed lines are same as in the left panels.
Filled squares indicate the estimated
redshifts of 15 observed GRBs.
In all panels, the photon index $\alpha$
is assumed to be 1.0 (thin lines) and 2.0 (thick lines).
Note that the $z_{\rm max}$ is significantly reduced when $\alpha=2.0$.
\label{fig1}}

\figcaption{Plots simlar to Fig. 1. Results for the total
sample for different opening angles in  the conic beam model. Opening
angles are indicated at the upper right corners in each panel.
In all panels, the photon index $\alpha$ is assumed to be 1.0 (thin lines)
and 2.0 (thick lines).
\label{fig2}}

\figcaption{Same plots as Fig. 2. for the long bursts.
In all panels, the photon index $\alpha$ is assumed to be 1.0 (thin lines)
and 2.0 (thick lines).
\label{fig3}}

\figcaption{Same plots as Fig. 2 for the short bursts.
In all panels, the photon index $\alpha$ is assumed to be 1.0 (thin lines)
and 2.0 (thick lines). 
\label{fig4}}

\figcaption{Same plots as Fig. 2 for the conic-beam case for the total sample.
We adopt a bi-Gaussian function as a number density function of GRB sources,
n(z), which is different from that used in Figs. 1-4.
 It gradually decreases at high redshifts.
The photon index $\alpha$ is assumed to be 1.0.
\label{fig5}}

\clearpage

\begin{thebibliography}{}
\bibitem[Blain and Natarajan (2000)]{bn00} Blain, A.W. \& Natarajan, P.
2000, \mnras, 312, 35

\bibitem[Blandford and K$\ddot{\rm o}$nigle (1979)]{bland79}
Blandford, R. D. \& K$\ddot{\rm o}$nigl, A. 1979, \apj, 232, 34

\bibitem[Chang and Yi (2000)]{cy00}
Chang, H.-Y. \& Yi, I. 2000, astro-ph/0005302

\bibitem[Che et al. (1999)]{che99}
Che, H., Yang, Y. \& Nemiroff, R. J. 1999, \apj, 516, 559

\bibitem[Djorgovski et al. (1998)]{dj98}
Djorgovski, S. G. et al. 1998, \apj, 508, L17

\bibitem[Graziani et al. (1999)]{graziani99}
Graziani, C., Lamb, D. Q., and Marion, G. H. 1999, A\&AS, 138, 469

\bibitem[Katz and Canel (1996)]{katz96}
Katz, J. I. \& Canel, L. M. 1996, \apj, 471, 915

\bibitem[Kim, Yang, and Yi (1999)]{kim99}
Kim, C., Yang, J., and Yi, I. 1999,
J. of Korean Phys. Soc., 34, 459

\bibitem[Kippen et al. (1998)]{kippen98}
Kippen, R. M. et al. 1998, \apj, 504, L27

\bibitem[Kouveliotou et al. (1993)]{ko93}
Kouveliotou, C. et al. 1993, \apj, 413, L101

\bibitem[Krumholz et al. (1998)]{krum98}
Krumholz, M. et al. 1998, \apj, 506, L81

\bibitem [Kulkarni et al. (1998)]{kul98}
Kulkarni, S. R. et al. 1998, Nature, 393, 35

\bibitem[Kumar(1999)]{ku99}
Kumar, P.  1999, \apjl, 523, L113

\bibitem[Kumar and Piran(2000)]{kupi99}
Kumar, P. and Piran, T. 2000, \apj, 535, 152

\bibitem[Lamb et al. (1993)]{lamb93}
Lamb, D. Q., Graziani, C., and Smith, I. A. 1993, \apj, 413, L11

\bibitem[MacFadyen and Woosley (1999)]{mac99}
MacFadyen, A. I. \& Woosley, S. E. 1999, \apj, 524, 262

\bibitem[Madau et al. (1998)]{madau98} Madau, P., Pozzetti,
L., and Dickinson, M. 1998, \apj, 498, 10

\bibitem[Mallozzi et al. (1996)]{mallozzi96} Mallozzi, R. S.,
Pendleton, G. N., and Paciesas, W. S. 1996, \apj, 471, 636

\bibitem[Mao, Narayan and Piran (1994)]{mao94}
Mao, S., Narayan, R. and Piran, T. 1994, \apj, 420, 171

\bibitem[Mao and Yi (1994)]{maoyi94}
Mao, S., \& Yi, I. 1994, \apj, 424, L131

\bibitem[Meegan et al.(1992)]{meegan92}
Meegan, C. A., et al.  1992, Nature, 355, 143

\bibitem[Metzger et al. (1997)]{metzger97}
Metzger, M. R. et al. 1997, Nature, 387, 878

\bibitem[Paciesas et al. (1999)]{batse4B}
Paciesas, W. S. et al. 1999, 1999, \apjs, 122, 465

\bibitem[Paczy$\acute{\rm n}$ski (1986)]{paczynski86}
Paczy$\acute{\rm n}$ski, B. 1986, \apj, 308, L51

\bibitem[Paczy$\acute{\rm n}$ski (1998)]{paczynski98}
Paczy$\acute{\rm n}$ski, B. 1998, \apj, 494, L45

\bibitem[Piran (1999)]{pi99}
Piran, T. 1999, Phys. Rep., 314, 575

\bibitem[Schmidt et al. (1988)]{schmidt88}
Schmidt, M., Higdon, J. C., and Hueter, G. 1988, \apj, 329, L85

\bibitem[Steidel et al. (1999)]{steidel99} Steidel, C., Adelberger,
K. L., Giavalisco, M., and Dickinson, M. 1999, \apj, 519, 1

\bibitem[Tavani (1998)]{tavani98} Tavani, M. 1998, \apj, 497, 915

\bibitem[Totani (1997)]{totani97}
Totani, T. 1997, \apj, 486, L71

\bibitem[Wang and Wheeler (1998)]{wang98}
Wang, L. \& Wheeler, J. C. 1998, \apj, 504, L87

\bibitem[Wijers et al. (1998)]{wijers98}
Wijers, R. A. M. J. 1998, \mnras, 294, L13

\bibitem[Woosley (1993)]{woo93}
Woosley, S. E. 1993, \apj, 405, 273

\bibitem[Woosley et al. (1999)]{woo99}
Woosley, S. E. et al. 1999, \apj, 516, 788

\bibitem[Yi (1993)]{yi93} Yi, I. 1993, \prd, 48, 4518

\bibitem[Yi (1994)]{yi94} Yi, I. 1994, \apj, 431, 543

\end{thebibliography}
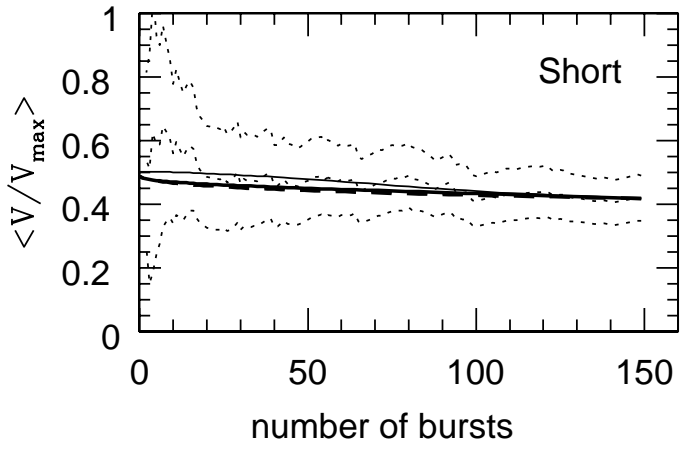
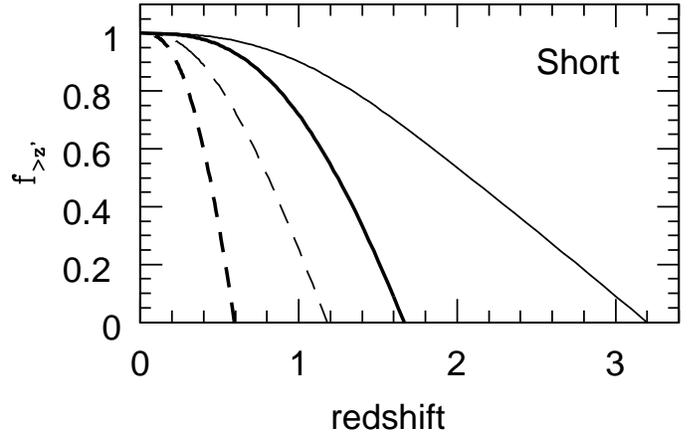
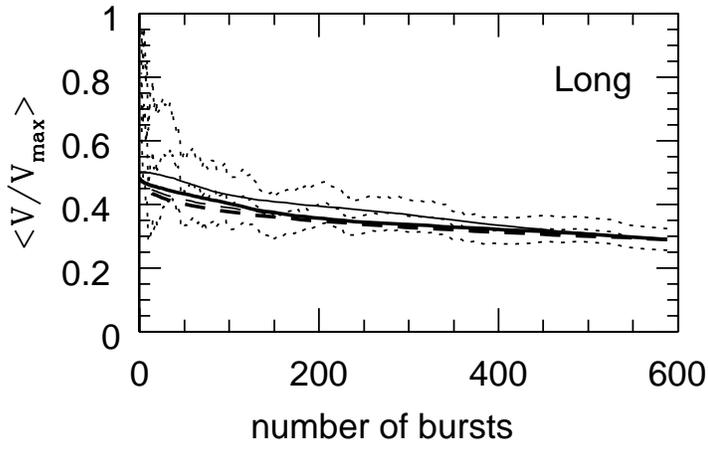
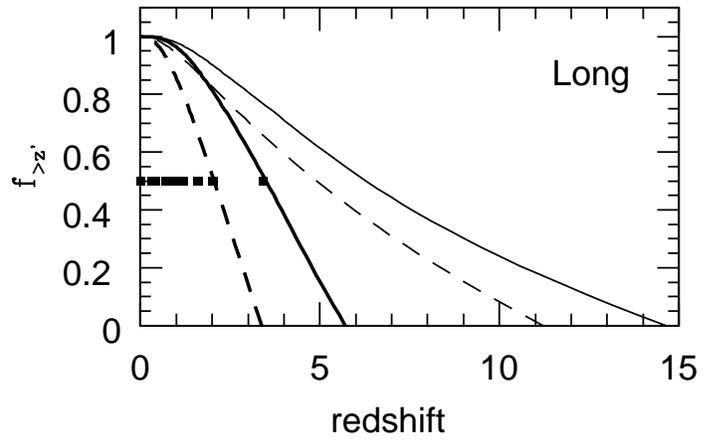
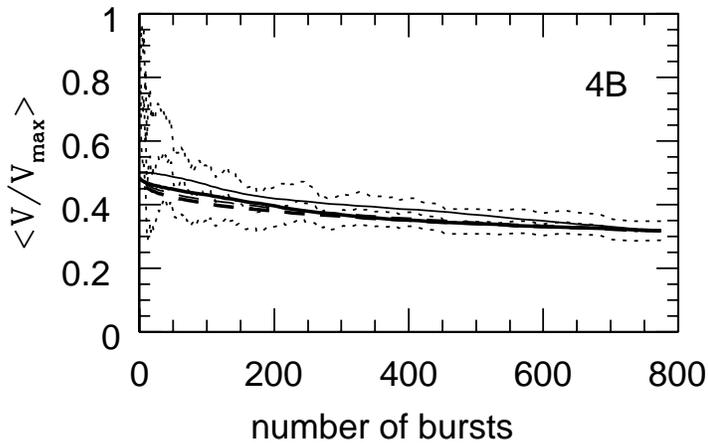
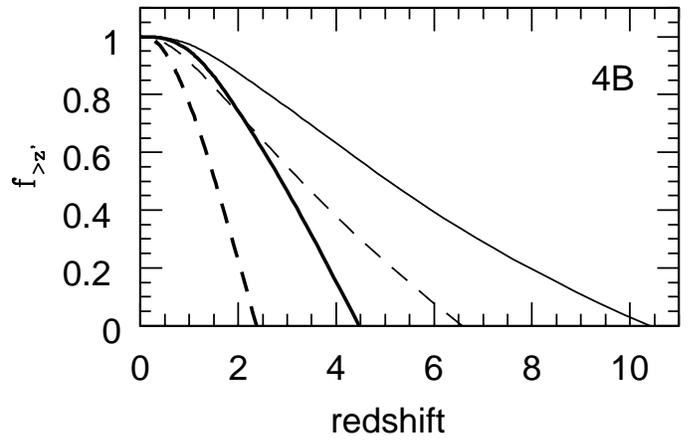
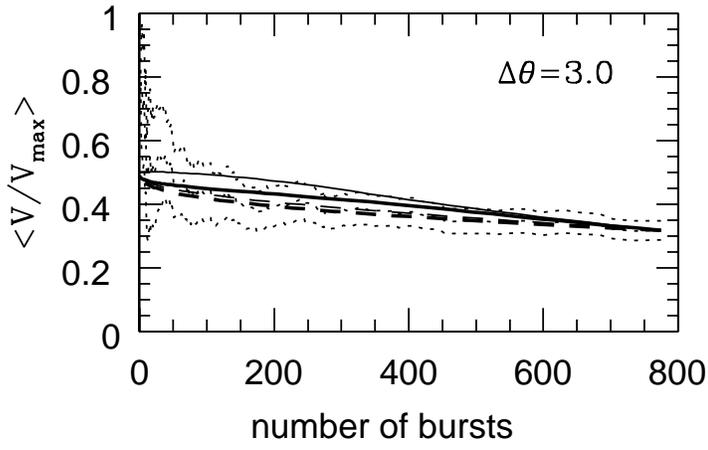
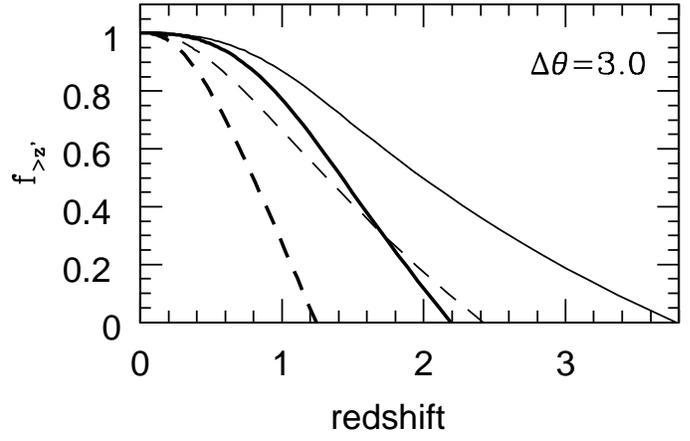
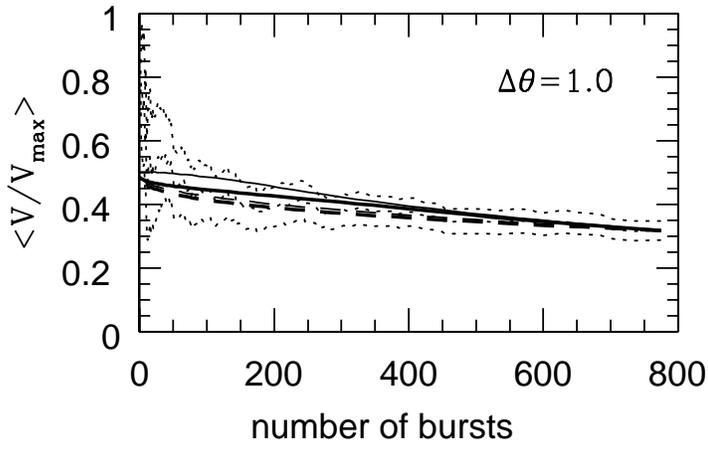
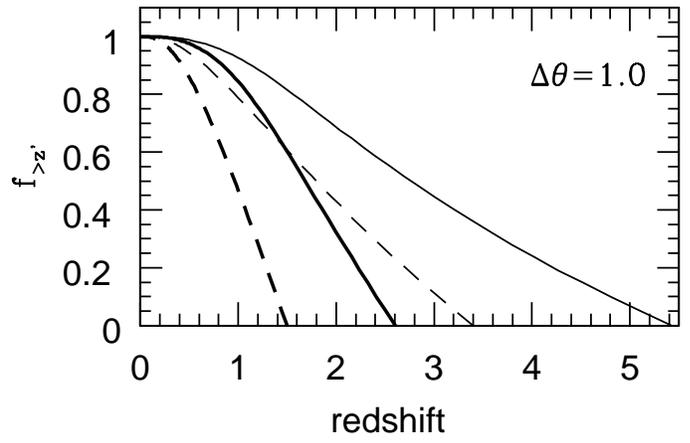
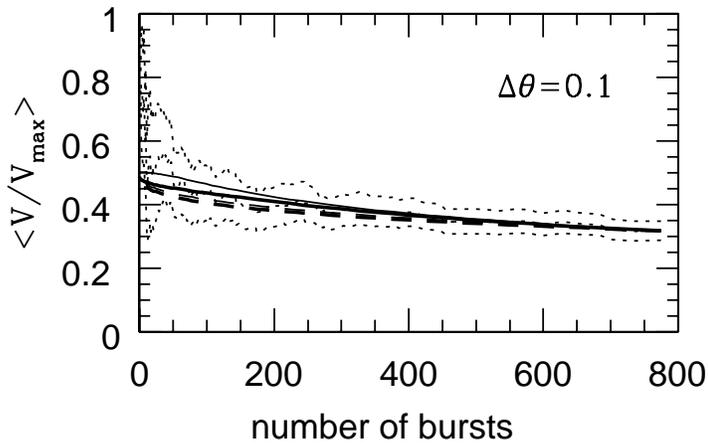
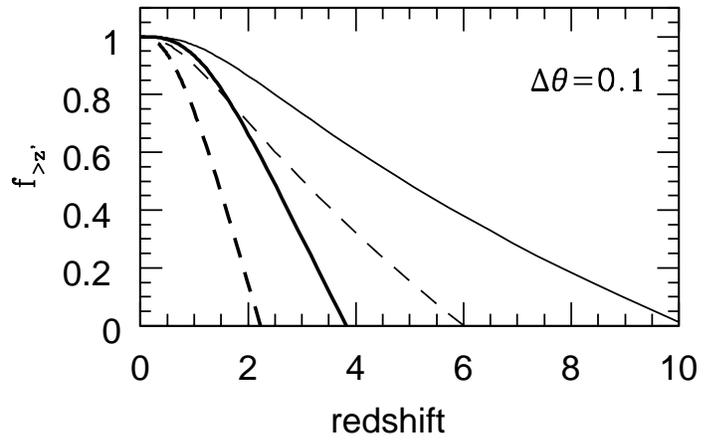
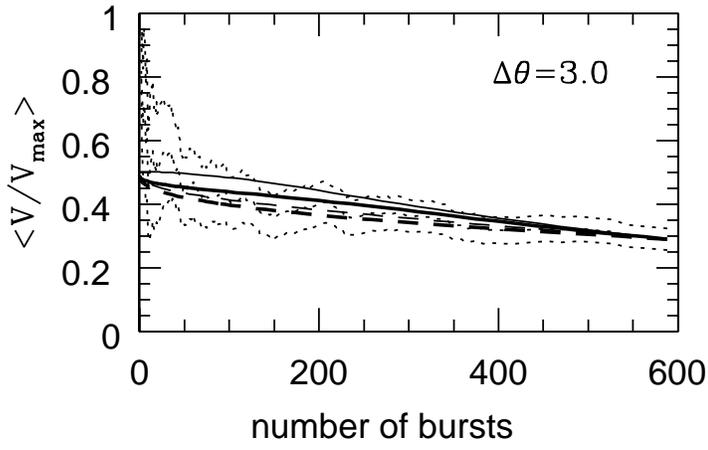
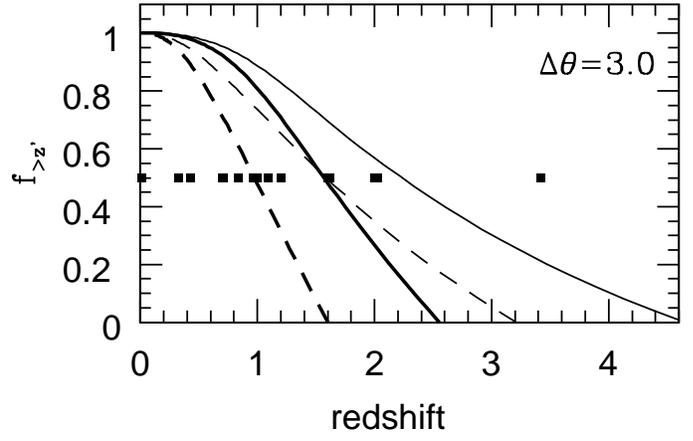
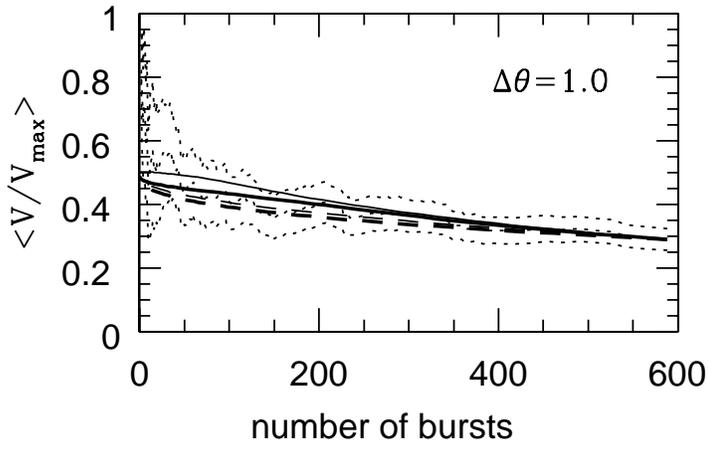
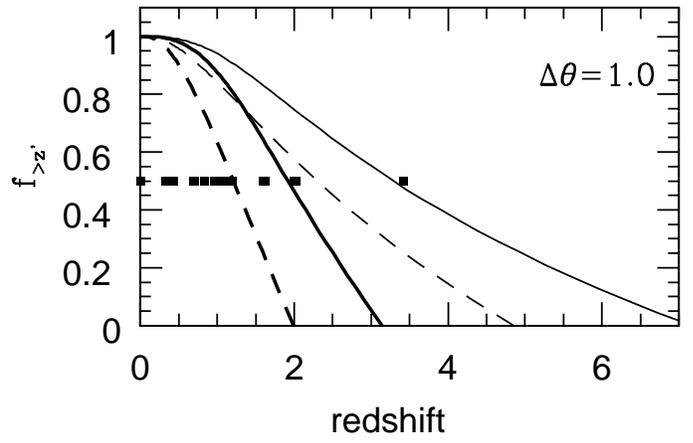
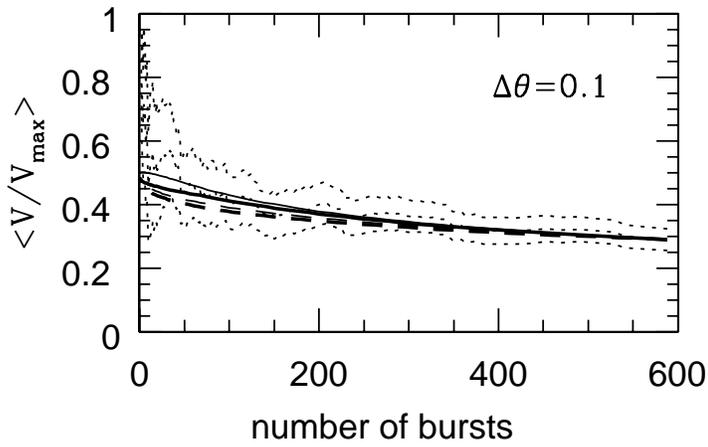
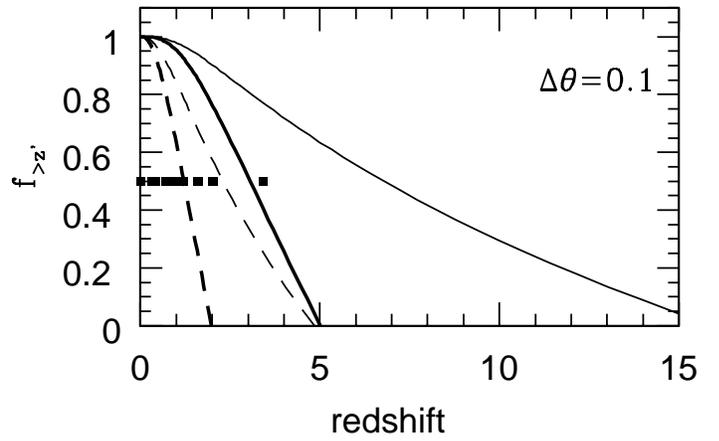
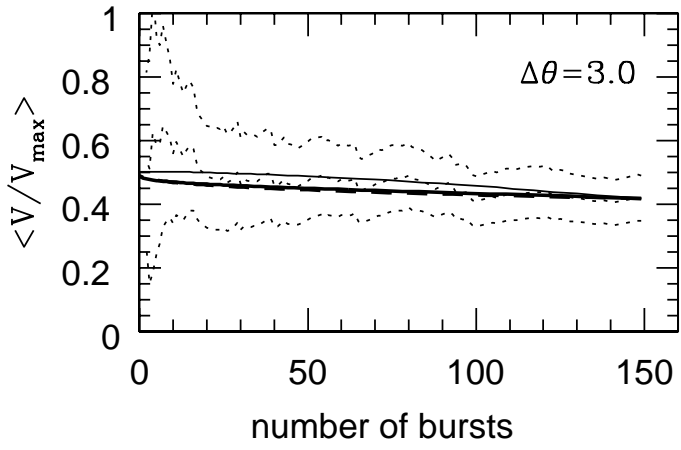
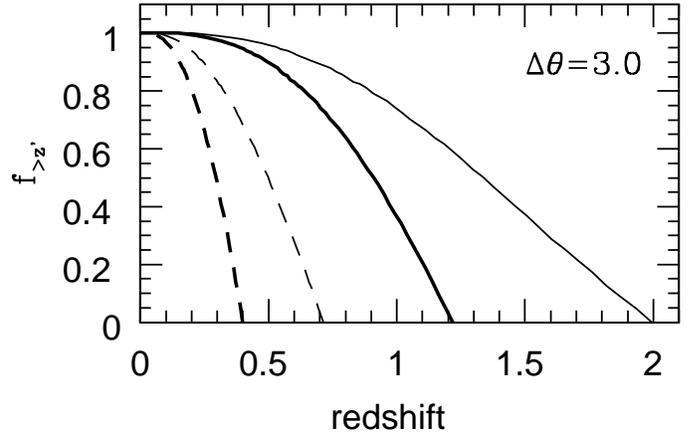
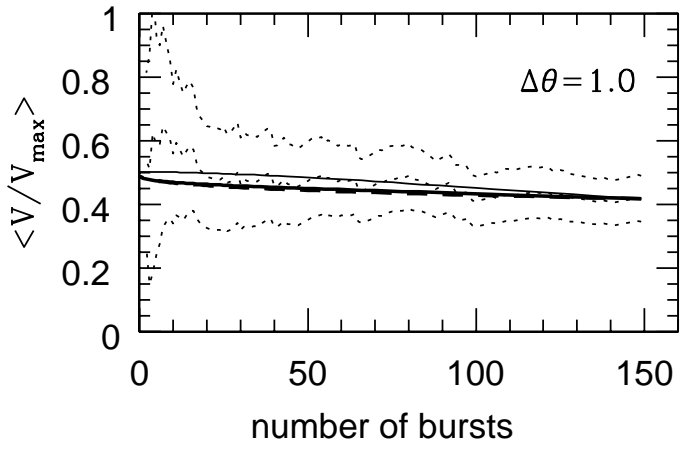
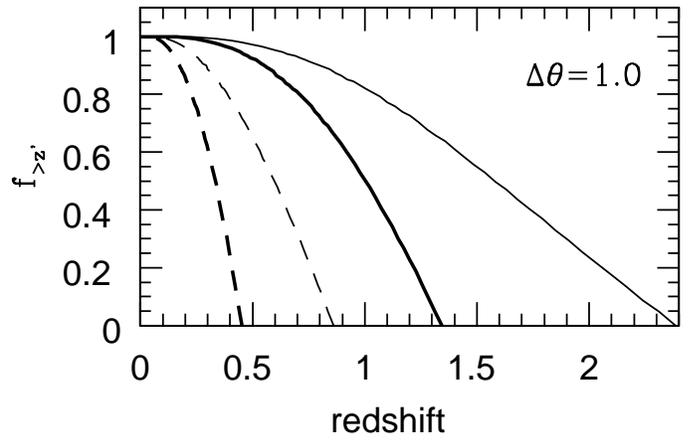
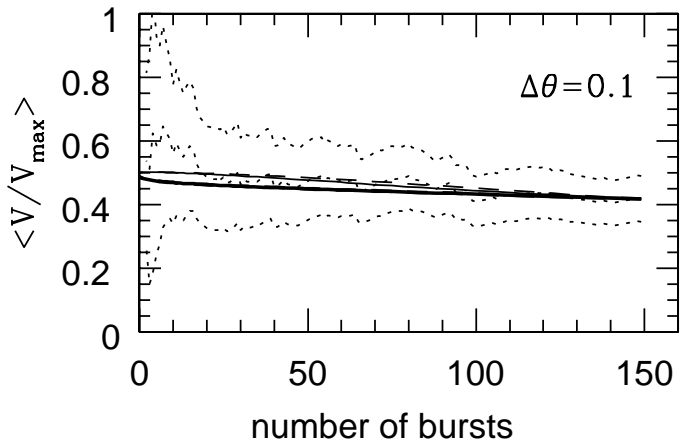
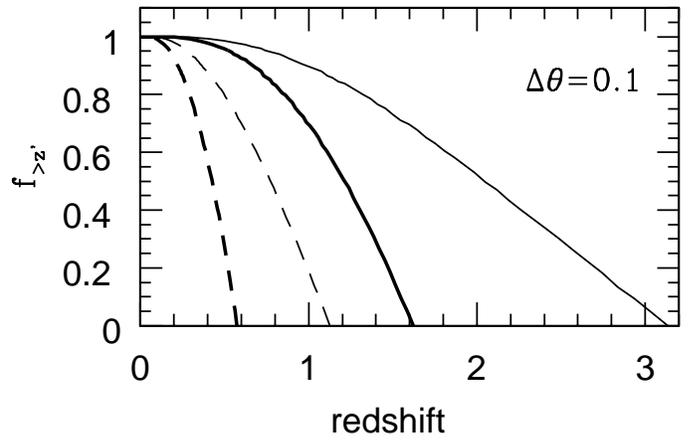
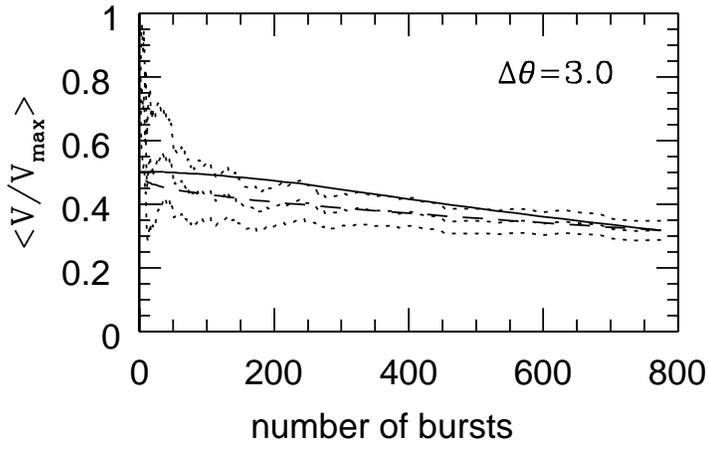
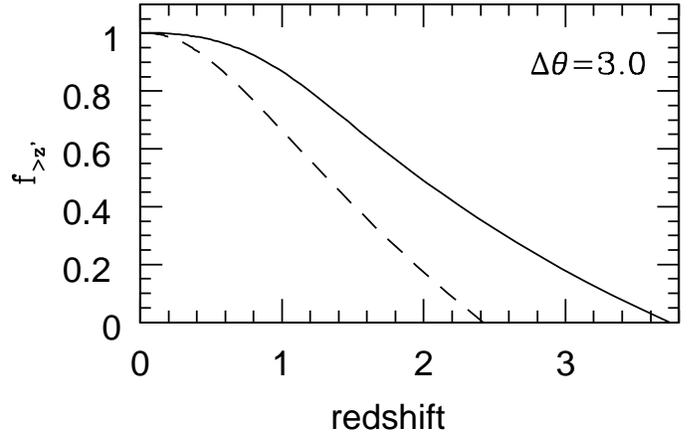
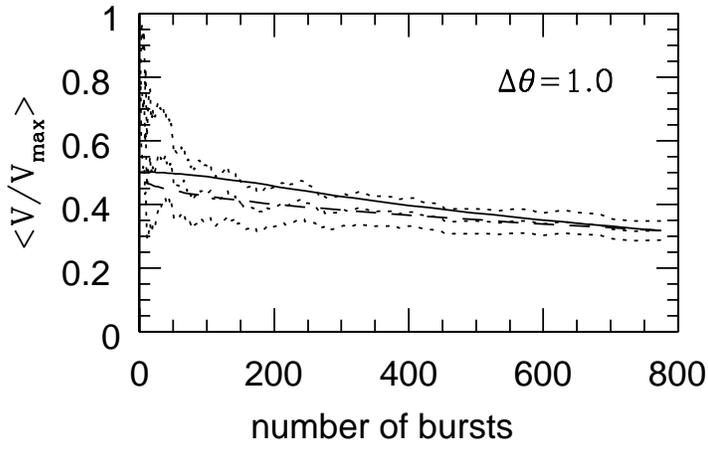
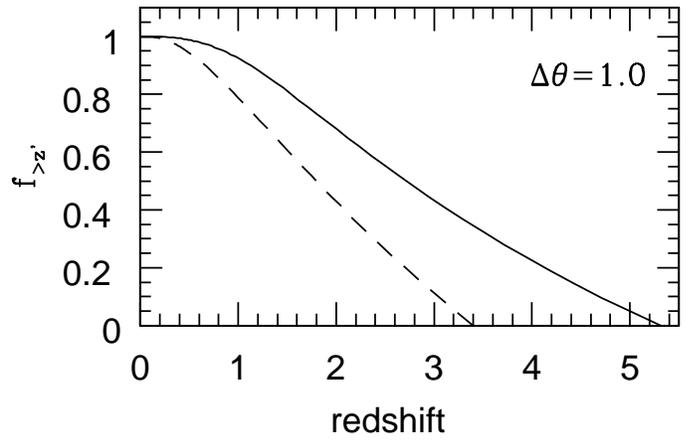
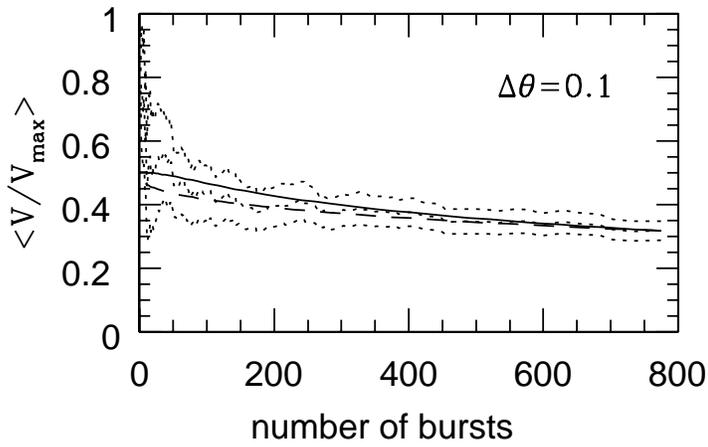
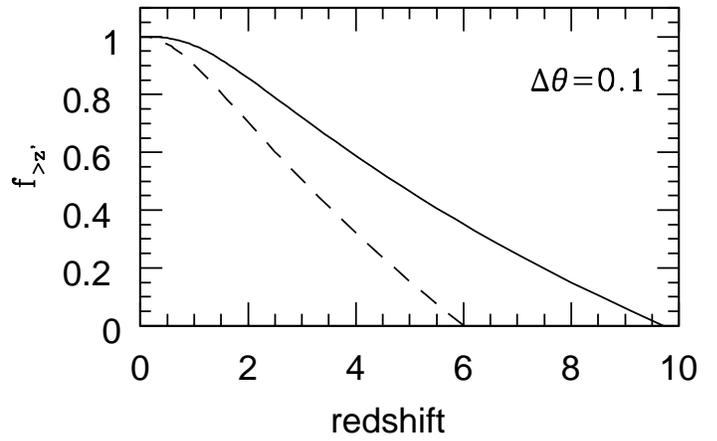
\end{document}